\documentstyle[prc,preprint,aps,epsfig,amssymb]{revtex}
\begin{document}
\title{Role of deformation in the nonmesonic decay of light hypernuclei}
\author{K. Hagino and A. Parre\~no} 
\address{Institute for Nuclear Theory, Department of Physics, \\
University of Washington, Seattle, WA 98195, USA}
\maketitle

\bigskip

\begin{abstract}

We discuss the nonmesonic decay of deformed p-shell hypernuclei. 
The Nilsson model with angular momentum projection is employed 
in order to take into account the deformation effects. 
The nonmesonic decay rate and the intrinsic $\Lambda$ asymmetry parameter 
decrease as a function of
the deformation parameter, while 
the ratio of the neutron- to 
proton-induced decay rates 
increases. 
We find that the deformation effects change these observables 
by about 10 \% for $^9_{\Lambda}$Be from the spherical limit.

\end{abstract}   


\section{Introduction}

One of the main issues of nuclear physics is to understand 
the nucleon-nucleon (NN) interaction. 
The $|\Delta S|=1$ NN interaction is particularly important 
in this respect, since 
the change of strangeness can be used as a signature to study 
both the parity-conserving (PC) and the parity-violating (PV) amplitudes. 
This is in clear contrast to the $\Delta S=0$ weak NN interaction,
where the weak PC signal is masked by the strong 
interaction. 

Due to the lack of stable $\Lambda$-particle beams, the weak decay of 
$\Lambda$-hypernuclei has been the only source of information
on the weak four-baryon $|\Delta S|=1$ interaction. 
Single $\Lambda$-hypernuclei are typically produced via either hadronic reactions, as 
($K^-_{stop}$,$\pi^0$)\cite{bnl} or ($\pi^+$,$K^+$)\cite{kek}, 
or electroproduction mechanisms,  
as (e,e' $K^+$)\cite{tjnaf}. 
These hypernuclei are typically produced in an excited state and 
reach 
their ground state by electromagnetic-$\gamma$ and/or particle emission. 
Once they are stable against strong decay, they decay via 
weak interaction mechanisms which are nonleptonic in nature and violate isospin,
parity and strangeness. 
Since the mesonic decay mode, $\Lambda \to \pi N$, is Pauli blocked 
in the nuclear medium, 
hypernuclei with $A \gtrsim 5$ predominantly decay through
the nonmesonic decay (NMD) mode, $\Lambda N \to NN$.

In order to learn about the weak $\Lambda N \to NN$ interaction 
from the theoretical side,
one has to take into account different inputs as accurately
as possible. 
These include the description of nuclear structure,
the choice of the strong BB potential model
\cite{nij99,juelich},
$\Delta I=1/2$ violations \cite{PRBM98} and
the importance of the 3N emission channel,
$\Lambda np \to nnp$ \cite{AP91,RVVO97}.
In Refs.\cite{PRB97,Go97}, a one-meson-exchange (OME) model was applied to 
calculate the nonmesonic decay observables of the $p$-shell
$^{11}_{\Lambda}$B and $^{12}_{\Lambda}$C
and the $s$-shell 
$^5_{\Lambda}$He and $^3_\Lambda$H hypernuclei. 
We included the virtual exchange of 
the ground-state pseudoscalar and vector mesons $\rho$, $\eta$,$\omega$,
$K$ and $K^*$, in addition to the 
long-ranged pion.
Except for the hypertriton, where the hypernuclear wave function was 
calculated exactly using the Faddeev formalism, the structure of the initial
hypernucleus was described in a shell-model framework which assumed
spherical configuration. In these calculations,
the strong baryon-baryon (BB) interaction was accounted for using the
Nijmegen BB potential model\cite{nij99}. Monopole form factors at
each vertex were included in order to regularize the weak potential,
while the weak baryon-baryon-meson coupling constants were
derived based on
SU$_{w}$(6) and soft-meson theorems.
The total NMD rate and the asymmetry in the distribution of emitted protons
from the decay of polarized hypernuclei were in good agreement with the 
experimental data.
However, the theoretical values for the neutron-to-proton ratio were found to
be very small compared to the experimental data. Several attempts have been
made to reconcile this discrepancy\cite{PRBM98,AP91,RVVO97,DFHT96,SS96,oka}, 
but none of them has solved this problem yet.

Our aim in this paper is to investigate how much these observables depend on 
the deformation of hypernuclei. 
All previous calculations were performed 
using the spherical configuration, 
however, it is well known that many $p$-shell nuclei are deformed in the 
ground state. For instance, the quadrupole deformation parameter extracted 
from the experimental quadrupole moment \cite{AS90} is $\beta_2=$ 0.65 
for $^{10}$B and $-0.71$ for $^{11}$C. 
It may be important to take these deformation effects into account in 
order to describe quantitatively the nonmesonic decay of $p$-shell 
hypernuclei. 
Deformed hypernuclei can be described using
several models such as the $\alpha$-cluster 
model \cite{bando} or the deformed self-consistent Hartree-Fock method. 
In fact, one can also use realistic wave functions obtained by 
a diagonalization of a shell-model Hamiltonian for p-shell nuclei, as
in Ref. \cite{cohen}. In the present paper, however, in order to perform a
systematic study, we use instead the Nilsson model\cite{GLN67,RS80} 
as a simplified Hartree-Fock 
method.

The paper is organized as follows. 
In Sec. II, we present the relevant formulae to evaluate the NMD observables
in a OME model.
In Sec. III, we briefly review the deformed 
shell model based on the Nilsson model. 
Sec. IV presents the deformation dependence of the nonmesonic observables
for the decay of $^{9}_{\Lambda}$Be,
whose $^8$Be core is known to be largely 
deformed. 
Although there is no experimental data for this hypernucleus at present, 
we choose this system as the simplest non-spherical p-shell hypernucleus
and as a representative example of deformed p-shell nuclei. 
We compare our theoretical predictions with the typical experimental
data for other p-shell hypernuclei.
Sec. V summarizes the paper. 

\section{Nonmesonic weak decay in a one-meson-exchange model}

Assuming that the initial hypernucleus is at rest,
the NMD rate is

\begin{equation}
\Gamma_{\rm nm} = \int  \frac{d^3 k_1}{(2\pi)^3}
 \int  \frac{d^3 k_2}{(2\pi)^3}
\sum_{^{M_I  \{R\} }_{\{1\} \{2\}}}
\,\, (2\pi) \,\, \delta(M_H-E_R-E_1-E_2)
\,\, \frac{1}{(2J+1)} \,\,
\mid {\cal M}_{fi} \mid^2  \ ,
\label{eq:rate0}
\end{equation}
where ${\cal M}_{fi}$ is the hypernuclear transition amplitude. 
The quantities
$M_H$, $E_R$, $E_1$ and $E_2$ are the 
mass of the hypernucleus,
the energy of the residual \mbox{$(A-2)$}-particle system, and the total
asymptotic energies of the emitted nucleons, respectively.
The integration variables ${\vec k}_1$ and ${\vec k}_2$ are 
the momenta of the two baryons in the final state.
The momentum conserving delta function has been used to
integrate over the momentum of the residual nucleus.
The sum, together with the factor $1/{(2J+1)}$, indicates an average
over the initial hypernucleus spin
projections, $M_I$, and a sum over all quantum numbers
of the residual
\mbox{$(A-2)$}-particle system, $\{R\}$, as well as the spin and isospin
projections of the exiting particles, $\{1\}$ and $\{2\}$.
In general, one can write
the total nonmesonic decay rate as
%
$\Gamma_{\rm nm}=\Gamma^{\Lambda N \to NN}=\Gamma_{\rm n}+\Gamma_{\rm p}$,
%
where $\Gamma_{\rm n}$ (\mbox{$\Lambda n \to nn$}) stands for the 
neutron-induced decay
and $\Gamma_{\rm p}$ (\mbox{$\Lambda p \to np$}) for the proton-induced one. 

In addition to the total and partial decay rates, 
we also calculate the intrinsic 
$\Lambda$ asymmetry parameter. 
When working with polarized hypernuclei
and in combination with coincidence measurements of the decay particles, 
one can study the angular distribution of particles coming from the
$\Lambda N \to NN$ weak decay. 
Due to the interference between the
PV and PC amplitudes, the distribution
of the emitted protons in the weak decay displays an angular asymmetry
with respect to the polarization axis.
The asymmetry ${\cal A}$, defined by
\begin{equation}
{\cal A}=P_y \frac{3}{J+1} \frac{Tr({\cal M}_{fi} S_y {{\cal M}_{fi}}^\dagger)}{
Tr({\cal M}_{fi} {{\cal M}_{fi}}^\dagger)} \, \ ,
\end{equation}
is expressed in terms of the hypernuclear polarization created
in the strong production reaction, $P_y$,
the $J$-spin operator along
the polarization axis, $S_y$, and the total spin of the initial hypernucleus,
$J$.
In Ref.\cite{RM92} it is shown that the asymmetry follows a simple
$\cos \chi$ dependence, {\it i.e.,} ${\cal A} = P_y A_p \cos \chi$, where
$\chi$ stands for the angle between the direction of the proton and the
polarization axis.
The hypernuclear
asymmetry parameter $A_p$ is characteristic of the hypernuclear weak 
decay process and
depends on $J$ and the intensity of protons exiting along the quantization axis
for the different spin projections of the hypernucleus.
At $\chi=0^\circ$, the asymmetry in the distribution of protons is thus
determined by the product ${\cal A} = P_y A_p$. 
In the following, we assume 
a weak coupling scheme 
where the $\Lambda$ hyperon is coupled only to the ground
state of the $(A-1)$-particle core. 
In this scheme,
simple angular momentum algebra relates the hypernuclear polarization
$P_y$ to the $\Lambda$ polarization $p_\Lambda$, and the hypernuclear asymmetry
parameter $A_p$ to the intrinsic $\Lambda$ asymetry parameter $a_\Lambda$, such
that ${\cal A} = p_\Lambda a_\Lambda = P_y A_p$ .

\begin{figure}[hbt]
\caption{Free Feynman diagrams for the $\Lambda N \to NN$ transition 
mediated by the
exchange of the nonstrange $\pi,\eta,\rho,\omega$ (left) and strange 
$K,K^*$ (right) mesons. 
The shaded circle (filled) stands for the weak (strong) vertex. 
}
\vspace*{0.5cm}
       \setlength{\unitlength}{1mm}
       \begin{picture}(100,180)
       \put(-70,-70){\epsfxsize=21cm \epsfbox{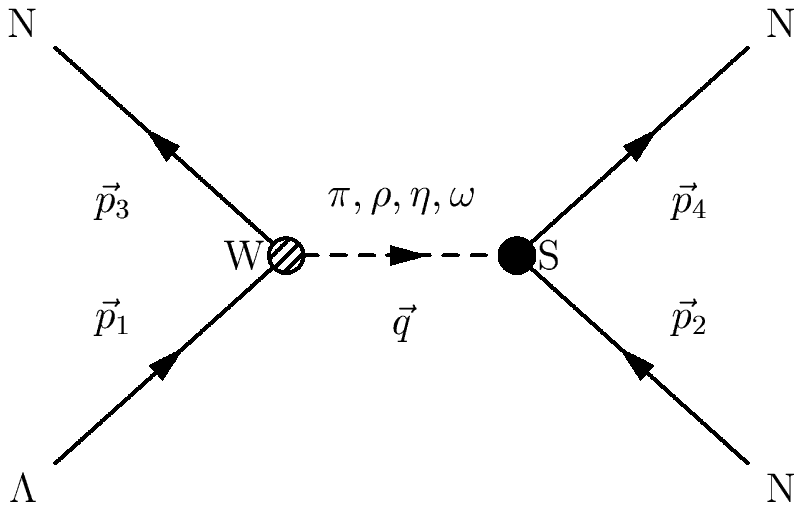}}
       \put(20,-70){\epsfxsize=21cm \epsfbox{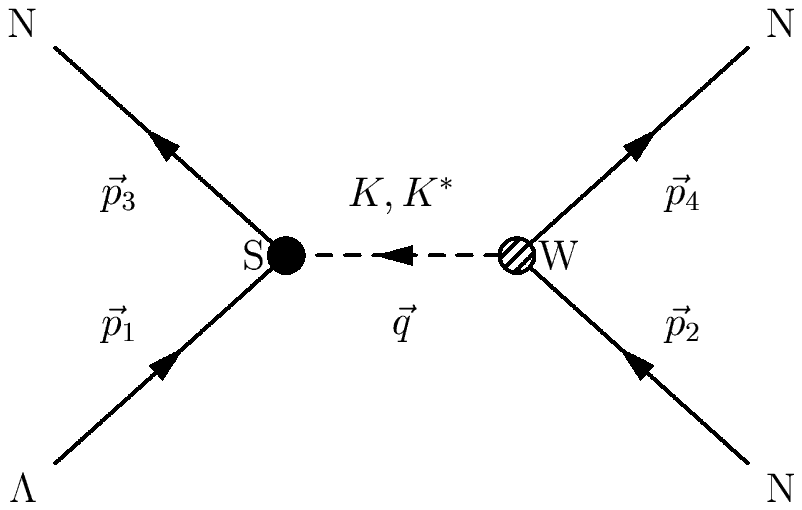}}
       \end{picture}
\label{fig1}
\end{figure}

The nonmesonic decay of hypernuclei proceeds through a two-body mechanism.
Therefore in order to evaluate the 
transition amplitude in Eq. (\ref{eq:rate0}), one has to decompose the (A-1)-core
wave function into a set of states in which a nucleon couples to the residual
(A-2)-particle state. This can be done using the 
Coefficients of Fractional Parentage (CFP), 
which are defined by 
\begin{equation}
|JM, TT_z\rangle = \sum_{J_R,T_R,j}\langle JT\{|J_RT_R,jt\rangle
\left[|J_RT_R\rangle \bigotimes |j t\rangle \right]_{JM,TT_z},
\end{equation}
where $J_R$ and $T_R$ are the spin and isospin of the residual nucleus.
The weak potential responsible for this transition 
can be obtained by making a nonrelativistic reduction 
of the free Feynman amplitude 
depicted in Fig. \ref{fig1}.
In Table \ref{tab:vert} we show the strong and weak vertices
for
pseudoscalar (PS) and vector (V)
mesons. $A, B, \alpha, \beta$ and $\epsilon$ stand for the appropriate 
baryon-baryon-meson weak coupling
constants, while $g$ ($g^{\rm V},g^{\rm T}$) represents the strong (vector,
tensor) coupling.
Details of the derivation of the transition potential can be found 
in Ref. \cite{PRB97} and here
only the final expression will be presented. For pseudoscalar mesons, the potential
is 
\begin{equation}
V_{ps}({\vec q}\,) = - G_F m_\pi^2
\frac{g}{2M} \left(
{\hat A} + \frac{{\hat B}}{2\overline{M}}
{\vec \sigma}_1 \, {\vec q} \,\right)
\frac{{\vec \sigma}_2 \, {\vec q}\, }{{\vec q}^{\; 2}+\mu^2} \ ,
\label{eq:pion}
\end{equation}
where $G_F {m_\pi}^2=2.21 \times 10^{-7}$ is the Fermi coupling constant,
${\vec q}$ is the momentum carried by the meson directed towards the
strong vertex, $\mu$ the meson mass and
$M$ ($\overline M$) is the average of the baryon masses at the
strong (weak) vertex (the other way around for the exchange of strange mesons).
For vector mesons, the potential is
\begin{eqnarray}
{V_{v}}({\vec q}\,)  &=&
G_F m_\pi^2
 \left( g^{\rm V} {\hat \alpha} - \frac{({\hat \alpha} + {\hat \beta} )
 ( g^{\rm V} + g^{\rm T} )} {4M \overline{M}}
({\vec \sigma}_1 \, \times {\vec q} \,)
({\vec \sigma}_2 \, \times {\vec q} \,) \right. \nonumber \\
& & \phantom { G_F m_\pi^2 A }
\left. - {\rm i} \frac{{\hat \varepsilon} ( g^{\rm V}+ g^{\rm T} )} {2M}
({\vec \sigma}_1 \, \times
{\vec \sigma}_2 \, ) {\vec q} \,\right)
\frac{1}{{\vec q}^{\; 2} + \mu^2} \ .
\label{eq:rhopot}
\end{eqnarray}
The values of the strong and weak couplings are listed in 
Table III of Ref. \cite{PRB97}.
In Eqs.~(\ref{eq:pion}) and (\ref{eq:rhopot}) the operators
${\hat A}$, ${\hat B}$, ${\hat \alpha}$, ${\hat \beta}$ and
${\hat \varepsilon}$ contain, apart from the weak coupling constants,
 the specific isospin dependence of the potential,
which is
${\vec \tau}_1 \cdot {\vec \tau}_2$ for the isovector $\pi$ and $\rho$ mesons,
${\hat 1}$ for the isoscalar
$\eta$ and $\omega$ mesons and
a combination of both operators for the isodoublet
$K$ and $K^*$.
In order to derive Eqs. (\ref{eq:pion}) and (\ref{eq:rhopot}) we assumed 
the validity of the 
$\Delta I=1/2$ rule, which is known to experimentally dominate the decay 
of $\Lambda$'s
into pions. $\Delta I=3/2$ transitions for vector mesons ($\rho$ and $K^*$) 
are easily accomodated\cite{PRBM98} 
in the formalism and the results we present here account for such
$\Delta I=1/2$ violations. 

We obtain a regularized potential by including a monopole form factor
at each vertex, $F({\vec q}^{\,\,2})= \displaystyle\frac{(\Lambda^2-\mu^2)}
{(\Lambda^2+{\vec q} ^{\,\,2})}$, where the value of the cut-off, $\Lambda$, 
different for each meson,
is taken from the J\"ulich hyperon-nucleon interaction\cite{juelich}.
To incorporate the effects of the strong NN interaction, 
we solve a T-matrix 
scattering equation in momentum space for the outgoing nucleons using
the Nijmegen\cite{nij99} potential models.
For the initial bound two-baryon system 
we use a spin 
independent parametrization
of the type
\begin{equation}
f(r)=\left( 1 - {\rm e}^{-r^2 / a^2} \right)^n + b r^2 {\rm
e}^{-r^2 / c^2} \ ,
\label{eq:cor}
\end{equation}
with $a=0.5$ fm, $b=0.25$ fm$^{-2}$, $c= 1.28$ fm, $n= 2$.
The results obtained with this parametrization\cite{sitges} lay in between of the ones
obtained with a microscopic finite-nucleus $G$-matrix calculation\cite{halder}
using the soft-core and hard-core Nijmegen\cite{nij89} models.

\section{Deformed shell model}

As we mentioned in the previous section, we use a weak coupling scheme for the
$\Lambda$ hyperon in the initial hypernucleus. 
To this end, we must describe the ground state of the core
nucleus which may be deformed.
The Nilsson model provides a simple and convenient framework to describe 
deformed nuclei, and has been widely used in the literature 
\cite{GLN67,RS80,HS95,SH97,K68,L80}. 
Its Hamiltonian 
consists of an anisotropic harmonic oscillator 
with the spin-orbit interaction as well as an angular momentum dependent 
term, which mocks up the deviation of the mean field potential from the 
harmonic oscillator
\begin{eqnarray}
H&=&H_0-\frac{4}{3}\sqrt{\frac{\pi}{5}} \, \delta \, m \, \omega_0^2 \, Y_{20}(\theta), \\
&=&-\frac{\hbar^2}{2m}\nabla^2
+\frac{1}{2} \, m \, \omega_0^2 \, r^2 + C \, \vec{l} \cdot \vec{s} + D \, (\vec{l \,\, }^2-
\langle \vec{l \,\, }^2 \rangle_N) 
-\frac{4}{3}\sqrt{\frac{\pi}{5}} \, \delta \, m \, \omega_0^2 \, Y_{20}(\theta). 
\label{Hnil}
\end{eqnarray}
Here $\delta$ is a deformation parameter, $\vec{l}$ and $\vec{s}$ are the single-particle 
orbital
and the spin angular momenta, and
$C$ and $D$ are adjustable parameters.
$\langle \vec{l\,}^2 \rangle_N=N(N+3)/2$ is the expectation value of
$\vec{l\,}^2$ averaged over one major shell with quantum number $N$.
The relation between $\delta$ and 
$\beta_2$ is given by \cite{ZZL91}
\begin{equation}
\beta_2=\frac{4}{3}\sqrt{\frac{\pi}{5}}
\frac{\delta}{1-2\delta/3} \, .
\end{equation}
Since the Nilsson Hamiltonian (\ref{Hnil}) violates rotational 
invariance, the total angular momentum $\vec{j}=\vec{l}+\vec{s}$ 
is not a good quantum 
number. However, the projection of $\vec{j}$ on to the $z$-direction, 
$k$, is conserved and the single-particle 
levels are characterized by $k$ and other quantum numbers. 
We expand a Nilsson single-particle level, $\psi_{k(q)}$, in terms of
the eigenfunctions of
the spherical harmonic oscillator Hamiltonian $H_0$, $\phi_{nljk}$, as 
\begin{equation}
\psi_{k(q)}=\sum_{nlj}x^{(q)}_{nljk} \phi_{nljk} \, , 
\end{equation}
where $q$ are quantum numbers other than $k$. 
We choose $x^{(q)}_{nlj-k}=(-)^{j-k}x^{(q)}_{nljk}$ so that the 
eigenvalues of the Nilsson Hamiltonian do not depend on the sign of 
the projection of total angular momentum \cite{L80}. 
We denote the creation operator of $\psi_k$ as $a^{\dagger}_k$ and 
that of $\phi_{jk}$ as $b^{\dagger}_{jk}$. We explicity express only 
the $j$ and $k$ quantum numbers to simplify the notation.
Intrinsic wave functions, i.e., eigenfunctions of the Nilsson Hamiltonian 
(\ref{Hnil}) are given by 
\begin{equation}
|\chi_K\rangle=a^{\dagger}_{k_1}a^{\dagger}_{k_2}\cdots 
a^{\dagger}_{k_n}|0\rangle 
=\prod^n_{i=1}\left(\sum_j x_{jk_i}b^{\dagger}_{jk_i}\right)|0\rangle, 
\label{intwf}
\end{equation}
where the $K$ quantum number is the sum over all $k_i$. 
The intrinsic wave function (\ref{intwf}) is not an eigenstate of the 
total angular momentum $\vec{J}$, and thus has to be projected out to a 
good angular momentum state. This can be achieved by using the projector
given
by \cite{RS80,HS95,SH97}
\begin{equation}
\hat{P}^J_{MK}=\frac{2J+1}{8\pi^2}\int d\Omega \,
D^J_{MK}(\Omega) \, \hat{R}(\Omega),
\end{equation}
where $\Omega$ are Euler angles, and 
$D^J_{MK}(\Omega)$ and $\hat{R}(\Omega)$ are the Wigner $D$-function and 
the rotation operator, respectively. 

For systems with a single-Nilsson level, such as $^8$Be which we discuss in
the next section, the CFP can be analitically 
obtained \cite{K68,L80}. 
Note that a single Nilsson level can accommodate up to four nucleons, 
i.e., two protons and two neutrons. 
For three particle systems ( 2 neutrons and 1 proton, for example), 
the wave function is given by 
\begin{eqnarray}
\Psi_{JM}&=&[N(3)_{J}]^{-1}\hat{P}^J_{M,K=k}a^{\dagger}_{k\nu}
a^{\dagger}_{-k\nu}a^{\dagger}_{k\pi}|0\rangle, \\
&=&[N(3)_{J}]^{-1}
\sum_{j_1,j_2,j_3}\sum_{J_{12}}
x_{j_1k}x_{j_2-k}x_{j_3k}\langle j_1kj_2-k|J_{12}0\rangle 
\langle J_{12}0 j_3k|Jk \rangle \nonumber \\
&&\times\left(\left[a^{\dagger}_{j_1\nu}a^{\dagger}_{j_2\nu}\right]_{J_{12}}
a^{\dagger}_{j_3\pi}\right)_{JM}|0\rangle, 
\label{wf3}
\end{eqnarray}
where $\pi$ stands for proton and $\nu$ for neutron. 
The normalization factor $N(3)_J$ is given \mbox{by \cite{K68}}
\begin{equation}
[N(3)_J]^2=\sum_{j,J_{12}}
\delta_{J_{12},even}(x_{jk})^2U(J_{12},k)
\langle J_{12}0jk|Jk\rangle^2,
\end{equation}
where 
\begin{equation}
U(J,k)=2\sum_{j_1,j_2}(x_{j_1k})^2(x_{j_2k})^2
\langle j_1kj_2-k|J0\rangle.
\end{equation}
The isospin of this system is 1/2. 
For four nucleon systems, the wave function reads 
\begin{eqnarray}
\Psi_{JM}&=&[N(4)_{J}]^{-1}\hat{P}^J_{M,K=0}a^{\dagger}_{k\nu}
a^{\dagger}_{-k\nu}a^{\dagger}_{k\pi}a^{\dagger}_{-k\pi}|0\rangle, \\
&=&[N(4)_{J}]^{-1}
\sum_{j_1,j_2}\sum_{j_3,j_4}\sum_{J_{12},J_R}
x_{j_1k}x_{j_2-k}x_{j_3k}x_{j_4-k}
\langle j_1kj_2-k|J_{12}0\rangle 
\langle J_{12}0 j_3k|J_Rk \rangle 
\langle J_Rk j_4-k|J0 \rangle 
\nonumber \\
&&\times\left(\left[\left(a^{\dagger}_{j_1\nu}a^{\dagger}_{j_2\nu}
\right)_{J_{12}}   
a^{\dagger}_{j_3\pi}\right]_{J_R}
a^{\dagger}_{j_4\pi}
\right)_{JM}|0\rangle, 
\label{wf4}
\end{eqnarray}

with the normalization given by \cite{K68}
\begin{equation}
[N(4)_J]^2=\sum_{J_{12},J_{34}}
\delta_{J_{12},even}\delta_{J_{34},even}
U(J_{12},k)U(J_{34},k)
\langle J_{12}0J_{34}0|J0\rangle^2. 
\end{equation}
The isospin of this wave function is 0. 
Comparing Eqs. (\ref{wf3}) and (\ref{wf4}), 
the CFP for the four particle 
system 
reads 
\begin{equation}
{\cal C} (j) = \langle JT\{|J_RT_R,jt\rangle
=-\sqrt{2}x_{j-k}
\langle J_Rkj-k|J0\rangle \frac{N(3)_{J_R}}{N(4)_J}. 
\label{cfp}
\end{equation}
%

\section{nonmesonic decay of $^9_{\Lambda}$Be}

Let us now apply the deformed shell model of Sec. III 
to the nonmesonic decay of $^9_{\Lambda}$Be.
The quadrupole moment of the neighbour nucleus $^9$Be was 
measured to be 5.86 $e$fm$^2$\cite{AS90}, 
from which we extract the quadrupole deformation 
parameter $\beta_2$=1.00 using the radius parameter 
$r_0$=1.2 fm. 
Several theoretical calculations suggest that the core nucleus $^8$Be
and the $^9_{\Lambda}$Be hypernucleus also have
similar deformation parameters with the same sign \cite{bando}. 
Our interest is to discuss such deformation effects on nonmesonic decay observables. 

As is discussed in Sec. II, the use of CFP allows us to write the hypernuclear transition 
amplitude ${\cal M}_{fi}$ in
terms of elementary two-body amplitudes. 
Therefore, our first task is to compute these coefficients for the 
core nucleus $^8$Be. 
We assume the inert spherical 
$^4$He core and explicitly work with only the four valence
nucleons. Diagonalizing the Nilsson Hamiltonian (\ref{Hnil}), one finds 
that the lowest Nilsson level for the valence nucleons 
has $k=1/2$ for prolate deformation \cite{RS80}. 
We diagonalize the Nilsson Hamiltonian in the $\Delta N $=0 states. 
Contributions from the $\Delta N=2$ can be neglected unless the deformation 
is large. 
The $k$=1/2 state is thus 
\begin{equation}
|\psi_{k=1/2}\rangle= x |\phi_{p_{3/2,1/2}}\rangle 
+ y |\phi_{p_{1/2,1/2}}\rangle ,
\end{equation}
where $x$ and $y$ are determined by diagonalizing the Nilsson 
Hamiltonian within this configuration space and depend upon the deformation of 
$^8$Be. 
Using Eq. (\ref{cfp}), the CFP's are found to be 
\begin{equation}
[{\cal C} (p_{3/2})]^2=\frac{3x^8+3x^6y^2+9x^4y^4} 
{3x^8+4x^6y^2+18x^4y^4+10y^8}
\end{equation}
for the p$_{3/2}$ state, and  
\begin{equation}
[{\cal C} (p_{1/2})]^2=\frac{x^6y^2+9x^4y^4+10y^8} 
{3x^8+4x^6y^2+18x^4y^4+10y^8} 
\end{equation}
for the p$_{1/2}$ state. Note that $[{\cal C} (p_{3/2})]^2+[{\cal C} (p_{1/2})]^2=1$. 
In the spherical limit, $x$=1 and $y=0$, so the CFP 
become $[{\cal C} (p_{3/2})]^2=1$ and $[{\cal C} (p_{1/2})]^2=0$. 
The CFP for the deeply bound 1s$_{1/2}$ state 
is just equal to 1 since $^4$He is a spin-isospin saturated nucleus. 

Our results for the nonmesonic decay rate, $\Gamma^{\rm nm}$, 
in units of the free $\Lambda$ decay rate, $\Gamma_\Lambda=3.8 \times 10^9 s^{-1}$,
the neutron-to-proton ratio, $\Gamma_{\rm n}/\Gamma_{\rm p}$,  
and the $\Lambda$ asymmetry parameter, $a_\Lambda$, are shown in Fig. 2 as a function of 
the deformation parameter $\beta_2$. 
We use an oscillator length $b_N$ of 1.65 fm for nucleons, 
so that 
the experimental root mean square radius of $^9$Be is reproduced. 
Following Refs. \cite{GLN67,RS80}, the parameters $C$ and $D$ in the 
Nilsson Hamiltonian (\ref{Hnil}) are 
taken to be $-0.16\hbar\omega_0$ and 0, respectively. 
As for the oscillator length $b_{\Lambda}$ for the 1s$_{1/2}$ wave function
of the $\Lambda$ hyperon, we estimate it to be 1.5 fm in order to 
reproduce its binding energy in $^9_\Lambda$Be ($=6.71 \pm 0.04$ MeV\cite{bando}). 
From the figure, we see that $\Gamma^{\rm nm}/\Gamma_\Lambda$ 
is 
a decreasing function of $\beta_2$, while $\Gamma_{\rm n}/\Gamma_{\rm p}$
and $a_\Lambda$ are increasing 
functions. As we have already mentioned, the deformation parameter of $^8$Be is 
expected to be close to 1. 
We notice that the nonmesonic decay observables are altered by about 10\% 
from the spherical limit at $\beta_2=1$. 

An important question is whether this effect is significant when comparing to
the experimental data. 
We note that the typical experimental uncertainties for nonmesonic decay 
of p-shell hypernuclei are: 7\% -- 17.5\% for the total decay rate 
\cite{B98,S91,N95}, 
46.2\% -- 84.2\% for the neutron-to-proton ratio\cite{S91,N95}, and 
50\% -- 1000 \% for the asymmetry \cite{A92}. 
These experimental uncertainties are much larger than the theoretical 
one originating from the deformation effects.
Thus we conclude that the spherical approximation gives a good estimate 
of the nonmesonic decay of p-shell nuclei, at least within the present
experimental precision. 

Before we close this section, we would like to stress that 
our conclusion is not altered qualitatively even if more realistic 
wave functions are used instead of the present schematic ones. 
For instance, using the
shell-model C.F.P. of 
Cohen and Kurath\cite{cohen} for the decay of
$^{12}_\Lambda$C,
we obtain 
$\Gamma^{\rm nm}/\Gamma_{\Lambda}=0.76$ and 
$\Gamma_{\rm n}/\Gamma_{\rm p}=0.22$. 
Those numbers have to be compared to the
spherical limit values of 
0.74 and 0.25, respectively.
As we see, the amount of deviation of those observables with respect to 
the spherical
limit is of the same order of the one given here, 
although their behaviour is opposite due partly to the fact that
the residual $^{11}$C
is oblate while $^8$Be is prolate.

\section{Summary}

We have discussed the role of nuclear structure in the nonmesonic 
decay of p-shell hypernuclei, especially focusing on the effects of 
deformation. To this end, we have used the Nilsson model with explicit 
angular momentum projection. We have studied the \mbox{nonmesonic} decay of 
$^9_{\Lambda}$Be as a typical example of deformed p-shell hypernuclei. 
We have shown that the deformation effects change the total NMD rate, 
the neutron-to-proton ratio and the 
$\Lambda$ asymmetry parameter 
by about 10 \% from the spherical limit. 
Although this value is not negligible, it still is smaller
than
the present typical experimental uncertainty and smaller than other 
theoretical uncertainties, e.g., the 
effects of SU(3) symmetry breaking\cite{PRB97,SS96} or $\Delta I=1/2$ 
violations\cite{PRBM98}. 
This indicates that 
the existing discrepancies 
between the experimental and theoretical values of 
hypernuclear weak decay observables cannot be attributed solely to deviations from the
spherical configuration,
and still remain an open question. New experiments are urged in
order to reduce the large experimental
error bars, which prevent any definite conclusion about the 
reliability of the theoretical
models.

Our conclusions may not be the same for heavier hypernuclei 
such as $^{238}_{\Lambda}$U\cite{A93,O97}. 
There are a lot of intruder states in such heavy deformed systems, unlike 
p-shell nuclei where there is only a few, or maybe 
zero, intruder states. 
Therefore, an interesting future work would be to discuss the nonmesonic 
decay of heavy hypernuclei including the deformation effects.  
For that purpose, 
the projected shell model developed in Refs. \cite{HS95,SH97}, 
which also uses the Nilsson model with angular momentum projection,
would provide a powerful tool to describe the structure of 
deformed hypernuclei.

\section*{Acknowledgements}
We thank David Brown and Amour Margarian for useful and illuminating discussions. 
This work was supported by
the U.S. Dept. of Energy under Grant No. 
DE-FG03-00-ER41132.

\newpage

\begin{figure}[hbt]
\caption{
The nonmesonic decay observables for $^9_\Lambda$Be as a function
of the deformation parameter $\beta_2$. }
\vspace*{-2.5cm}
       \setlength{\unitlength}{1mm}
       \begin{picture}(100,180)
       \put(0,-25){\epsfxsize=13cm \epsfbox{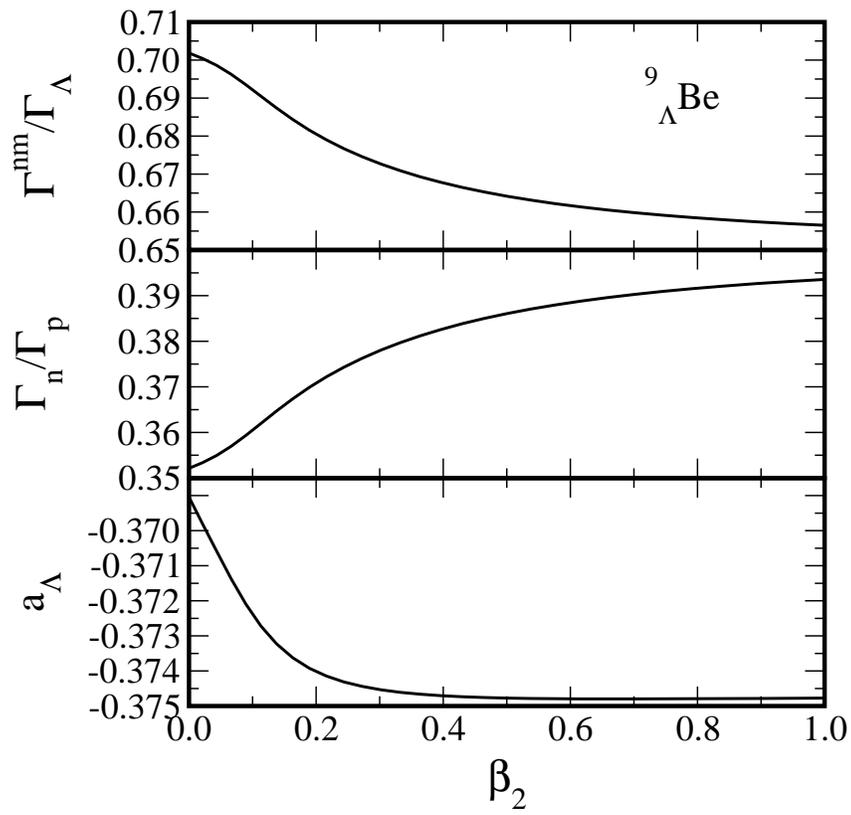}}
       \end{picture}
\end{figure}

\begin{table}[hbt]
\caption[]{Vertices entering the expression of the Feynman amplitude of
Fig. \ref{fig1} for pseudoscalar (PS) and vector (V) mesons. The weak vertices
are in units of $G_F {m_\pi}^2$.}
\vspace*{-2pt}
\begin{center}
\def\arraystretch{1.5}
\begin{tabular}{lcccc}
\hline
 &\phantom{merda} & PS &\phantom{merda} & V \\
\hline
Strong &\phantom{merda} &  ${\rm i} g \gamma_5$ &\phantom{merda} &
$g^{\rm V} \gamma^\mu + {\rm i}
\displaystyle\frac{ g^{\rm T}}{2M}\sigma^{\mu \nu} q_\nu$ \\
\hline
Weak &\phantom{merda} & ${\rm i} (A + B \gamma_5)$ &\phantom{merda} &
$\alpha \gamma^\mu - \beta {\rm i}
\displaystyle\frac{\sigma^{\mu \nu} q_\nu}
{2 \overline{M}}+\varepsilon\gamma^\mu \gamma_5$ \\
\hline
\end{tabular}
\end{center}
\label{tab:vert}
\end{table}

\end{document}